%% file: main.tex
\documentclass{IEEEtran}

\input{defs.tex}
\usepackage[group-separator={,}]{siunitx}
\usepackage{caption}
\usepackage{amsmath}
\usepackage{graphicx}
\usepackage{subcaption}
\usepackage{comment}
\usepackage{fancyhdr}
\newtheorem{theorem}{Theorem}

\usepackage{url}
\usepackage[
    margin=0.7in,
    bottom=0.75in,
]{geometry}

\begin{document}
\pagenumbering{arabic}

%
\title{
Fast Grid Emissions Sensitivities using Parallel Decentralized Implicit Differentiation
}

\author{
\IEEEauthorblockN{
Anthony Degleris$^{*,1}$, 
Lucas Fuentes Valenzuela$^{*, 1}$, 
Ram Rajagopal$^{1,2}$, 
Marco Pavone$^3$, 
Abbas El Gamal$^1$\\}
\IEEEauthorblockA{1. Department of Electrical Engineering\\ 2. Department of Civil \& Environmental Engineering\\ 3. Department of Aeronautics and Astronautics \\
Stanford University, Stanford, California\\
* equal contribution}
}


\maketitle

\input{body/0.Abstract}

\begin{IEEEkeywords}  
Sensitivity analysis,
distributed optimization,
implicit differentiation,
locational marginal emissions.
\end{IEEEkeywords}

\input{body/1.Introduction}
\input{body/2.LMEs}
\input{body/3.DID}

\input{body/4.NumericalExperiments}
\input{body/5.Conclusion}

\section{Acknowledgements}

Anthony Degleris is supported by the U.S.\ Department of Energy, Office
of Science, Office of Advanced Scientific Computing Research, Department
of Energy Computational Science Graduate Fellowship under Award Number
DE-SC0021110.

This research used resources of the National Energy Research Scientific Computing Center (NERSC), a U.S.\ Department of Energy Office of Science User Facility located at Lawrence Berkeley National Laboratory, operated under Contract No. DE-AC02-05CH11231 using NERSC award DDR-ERCAP0026889.



\bibliographystyle{IEEEtran}
%

\bibliography{paperpile}

\end{document}

%% file: defs.tex
\newcommand{\reals}{\mathbf{R}}

\newcommand{\diag}{\mathbf{diag}}

\newcommand{\initial}{\textrm{init}}
\newcommand{\final}{\textrm{final}}
\newcommand{\vectorize}{\mathbf{vec}}
\newcommand{\snext}{s^\mathrm{next}}

%% file: body/0.Abstract.tex
\begin{abstract}  
Marginal emissions rates---the sensitivity of carbon emissions to electricity demand---are important for evaluating the impact of emissions mitigation measures.
Like locational marginal prices, \textit{locational} marginal emissions rates (LMEs) can vary geographically, even between nearby locations, and may be coupled across time periods because of, for example, storage and ramping constraints.
This temporal coupling makes computing LMEs computationally expensive for large electricity networks with high storage and renewable penetrations.
Recent work demonstrates that decentralized algorithms can mitigate this problem by decoupling timesteps during differentiation.
Unfortunately, we show these potential speedups are negated by the sparse structure inherent in power systems problems.
We address these limitations by introducing a parallel, reverse-mode decentralized differentiation scheme that never explicitly instantiates the solution map Jacobian.
We show both theoretically and empirically that parallelization is necessary to achieve non-trivial speedups when computing grid emissions sensitivities.
Numerical results on a 500 node system indicate that our method can achieve greater than 10x speedups over centralized and serial decentralized approaches.
%
\end{abstract}

%% file: body/1.Introduction.tex
\section{Introduction}


In order to rapidly decarbonize the grid, decision makers need to identify and understand the best opportunities for reducing carbon emissions. 
Specifically, the effect of policies that shift electricity demand or supply are highly dependent on the emissions rates of the \textit{marginal} generators, i.e. those that are directly impacted.
\textit{Marginal emissions rates}, the sensitivities of emissions with respect to electricity consumption, are key metrics to evaluating these electricity-shifting policies~\cite{Siler-Evans2012-ve,Rudkevich2012-tx, Holland2022-sz}. 
These metrics have been used to develop operational strategies for server farms~\cite{Lindberg2021-ul} and quantify the impact of increasing electric vehicle penetration~\cite{Powell2022-ld}, among other applications.

At the regional level, e.g., aggregated across a balancing authority, emissions and demand data are often widely available, and marginal emissions rates can be estimated accurately using simple data-driven regression methods~\cite{Deetjen2019-ok}.
However, these methods do not distinguish between different nodes in the network and therefore ignore the non-negligible effects of transmission constraints.
In addition,
the time-varying nature of renewable generation has placed a greater emphasis on generator ramping constraints and led to a large increase in grid-scale battery storage~\cite{US_Energy_Information_Administration_EIA_undated-ek}, effectively coupling dispatch decisions across different time periods. 
These different sources of temporal coupling make incorporating dynamic constraints essential to obtaining accurate marginal emissions estimates in systems with high renewable and storage penetrations.
\textit{Dynamic, locational} marginal emissions (dynamic LMEs) address these limitations by giving marginal emissions rates at the nodal level and accounting for temporal constraints.
Because of the size and underlying complexity of modern electricity systems, dynamic LMEs are difficult to estimate using data-driven methods.
Instead, they can be computed exactly using the system operator's dispatch model~\cite{Ruiz2010-nj}.
In particular, recent work shows how to compute dynamic LMEs by differentiating through convex optimization-based dispatch problems~\cite{Fuentes_Valenzuela2023-dj}.

Unfortunately, this same temporal coupling greatly increases the cost of computing LMEs because time periods depend on one another.
In the worst case, the complexity of differentiating through optimization problems is cubic in the problem size. 
Computing dynamic LMEs in real time and giving accurate signals for emissions-driven operational strategies can thus be time-prohibitive, especially for large networks with high temporal resolution.
Recent work~\cite{Fuentes_Valenzuela2024-qp} addresses this computational issue by introducing a decentralized technique for implicit differentiation.
At a high level, the method decomposes the problem at each time period into a local component and a correction term that accounts for the coupling constraints---such as battery dynamics and ramping limits---between different time periods.
In power systems, however, problems are inherently sparse, greatly reducing the complexity of solving linear systems and potentially limiting the speedups of the method in~\cite{Fuentes_Valenzuela2024-qp}.

\paragraph*{Contribution}
In this work, we address this limitation and propose an efficient decentralized framework for computing dynamic LMEs in sparse power systems problems.
In particular, we introduce a parallel, reverse-mode variant of decentralized implicit differentiation and use it to efficiently compute LMEs.
Our method never explicitly instantiates the solution map Jacobian (forward-mode differentiation) and instead uses \textit{vector-Jacobian products}, i.e., reverse-mode differentiation.
We show that both centralized and decentralized implementations of reverse-mode differentiation are significantly faster than their corresponding forward-mode implementations for problems in power systems.
When comparing both reverse-mode implementations, our theoretical and empirical results indicate that parallelization is necessary to achieve non-trivial speedups for sparse systems.
We then show that our parallel decentralized algorithm significantly accelerates LME computation compared to both centralized and forward-mode decentralized approaches, achieving more than a 10x speedup in medium-sized cases.


\subsection{Notation}

We denote $\reals$ the set of real numbers, $\reals^N$ the set of $N$-dimensional real vectors, and $\reals^{N \times M}$ the set of $N \times M$ matrices with real-valued entries. 
The entry in the $n$th row and $m$th column of a matrix $M$ is denoted $M_{nm}$.
However, when confusion would otherwise arise, we use comma separated indices $M_{n, m}$.
Consider a multivariate function $F(x) : \reals^M \rightarrow \reals^N$.
We let $DF(x) \in \reals^{N \times M}$ denote the Jacobian matrix of $F$ evaluated at $x$; the entry $DF_{nm}(x)$ is the derivative of $F_n$ with respect to $x_m$.
For functions with multiple arguments $F(x, y) : \reals^N \times \reals^L \rightarrow \reals^L$, we use $\partial_1 F(x, y) \in \reals^{L \times N}$ and $\partial_2 F(x, y) \in \reals^{L \times L}$ to denote the partial Jacobians of $F$ with respect to its first and second arguments, respectively.
Sometimes we will also use $\partial_x F(x, y)$ and $\partial_y F(x, y)$ when the argument is clear and keeping track of indices would be otherwise difficult.
Finally, for matrix-valued functions $F(x) : \reals^N \rightarrow \reals^{L \times L}$, we will use $DF(x) \in \reals^{L^2 \times L}$ to refer to the Jacobian of the vectorized function $\vectorize(F(x))$, unless the context makes it such that this introduces an ambiguity.

%% file: body/2.LMEs.tex
\section{Locational Marginal Emissions}\label{sec:LMEs}

In this section, we introduce a dynamic electricity dispatch model that incorporates load, generation, transmission, and storage---similar to those solved by major system operators.\footnote{
We exclude other type of constraints, such as generator ramping constraints, for simplicity. 
However, the methodology applies directly to any type of intertemporal constraint.
}
Then, we define locational marginal emissions rates mathematically.
Finally, we explain a generic method for calculating them from~\cite{Fuentes_Valenzuela2023-dj} and describe its computational complexity.

\subsection{Dispatch Model}

We consider a (DC-linearized) optimal power flow problem on a network with $N$ nodes over $T$ time periods.
At time $t$, every node $n$ on the network has fixed demand for power $d_{nt} \in \reals_+$ and a generator that produces power $g_{nt} \in \reals_+$.
Each generator $n$ incurs a cost $c_{nt} g_{nt}$ to produce $g_{nt}$ units of power at time $t$.
The generator power outputs are constrained by their maximum outputs, $0 \leq g_{nt} \leq g_{nt}^{\max}$, where $g^{\max} \in \reals_+^{N \times T}$.

The network also has $K$ batteries that discharge power into the network $p_{kt} \in \reals$ where, like a generator, $p_{kt} > 0$ is a power injection (discharging) into the network, and $p_{kt} < 0$ is a withdrawal (charging).
The batteries have power constraints $-p_k^{\max} \leq p_{nt} \leq p_k^{\max}$, where $p^{\max} \in \reals_+^{K}$.
To keep the notation light, we assume the batteries are lossless\footnote{The methodology and results laid out in this paper remain valid if batteries are not 100\% efficient.}, so the battery state of charge at time $t+1$ is $s_{k, t+1} = s_{kt} - p_{kt}$, where $s \in \reals_+^{K \times (T+1)}$.
The state of charge is constrained to lie between $0 \leq s_{kt} \leq s^{\max}_k$, where $s^{\max} \in \reals_+^K$, and has initial and final conditions $s_{k,1} = s_k^{\initial}$ and $s_{k, T+1} = s_k^{\final}$, respectively.
Each battery $k$ is connected to a node $i(k)$; this is represented by the node-battery incidence matrix $B \in \reals^{N \times K}$, where $B_{nk} = 1$ if $n = i(k)$ and $B_{nk} = 0$ otherwise.

Lastly, the network has $M$ transmission lines.
Each transmission line $m$ has flow $f_{mt}$ at time $t$, constrained by the capacity constraint $-f_m^{\max} \leq f_{mt} \leq f_m^{\max}$, where $f_m^{max} \in \reals_+^{M}$.
Line $m$ is connected to two nodes $\ell_1(m)$ and $\ell_2(m)$.
All the line connections are represented by the node-branch incidence matrix $A \in \reals^{N \times M}$, which has entries
\begin{align*}
    A_{nm} = \begin{cases}
        1 & n = \ell_1(m) \\
        -1 & n = \ell_2(m) \\
        0 & \textrm{otherwise}.
    \end{cases}
\end{align*}
The flows on the transmission lines are also constrained by Kirchhoff's law, $f = \diag(b) A^T \theta$, where $\theta \in \reals^{N \times T}$ is a matrix of nodal voltage phase angles over time and $b \in \reals^M$ is the vector of branch susceptances.
At the reference node 1, we fix the voltage phase angle to zero, i.e., $\theta_{1,t} = 0$.
Finally, the network must satisfy the power balance constraint, $g + p - d = Af$.
The \textit{dispatch problem} is the optimization problem,
\begin{align} \label{eq:dispatch}
\begin{array}{lll}
    \textrm{minimize} 
    & \sum_{n, t} c_{nt} g_{nt} \\[0.5em]
    \textrm{subject to} 
    & g + Bp - d = Af \\
    & f = \diag(b) A^T \theta \\
    & 0 \leq g \leq g^{\max} \\
    & -f^{\max} \leq f \leq f^{\max} \\
    & -p^{\max} \leq p \leq p^{\max} \\
    & 0 \leq s \leq s^{\max} \\
    & \theta_{1,t} = 0, \quad &\forall t, \\
    & s_{k,1} = s_k^{\initial} & \forall k, \\
    & s_{k, T+1} = s_k^{\final} & \forall k \\
    & s_{k, t+1} = s_{kt} - p_{kt}, \quad &\forall k, t \\
\end{array}
\end{align}
where the variables are $g, \theta \in \reals^{N \times T}$, $f \in \reals^{M \times T}$, $p \in \reals^{K \times T}$, and $s \in \reals^{K \times (T+1)}$.
Although we use~\eqref{eq:dispatch} in all our derivations going forward, our results apply to broader classes of dispatch models, such as models with multiple (or zero) generators at each node, batteries with losses, non-linear costs, or generator ramping limits;
in general, our method in Section~\ref{sec:DID} works for broad range of convex optimization-based dispatch models and does not depend on the specifics of the problem formulation; refer to~\cite{Fuentes_Valenzuela2024-qp} for details.

\subsection{Marginal Emissions}

Let $\mathcal D \subset \reals^{NT}$ be the set of demand vectors $d$ such that~\eqref{eq:dispatch} is feasible and has a unique solution.
We define the \textit{solution map} $g^*(d) : \mathcal D \rightarrow \reals^{N \times T}$ to be the function that maps a demand vector $d$ to the optimal choice of generation schedule $g$.
This function is well-defined if the solution to~\eqref{eq:dispatch} exists and is unique, which can be guaranteed, for example, by adding a negligible quadratic penalty to each variable.
The total emissions across the system is given by $E(g) = \sum_{n, t} e_{nt} g_{nt}$, where $e_{nt} \in \reals$ is the emissions rate of generator $n$ at time $t$.
Then, total emissions as a function of demand is $H(d) = E(g^*(d))$.
The \textit{locational marginal emissions rates} (LMEs) are the sensitivities of total emissions with respect to demand, i.e., the gradient
\begin{align}
    \lambda(d) := \nabla H(d).
\end{align}
The goal of this paper is to give an efficient algorithm for computing $\lambda(d)$.

\subsection{Solution via Implicit Differentiation}
By the chain rule, we know that
\begin{align}
    \nabla H(d) = Dg^*(d)^T \vectorize(e),
    \label{eq:LMEs_implicit}
\end{align}
where $Dg^*(d) \in \reals^{NT \times NT}$ is the Jacobian of $g^*$.
Therefore, calculating the LMEs requires differentiating the optimal variable of an optimization problem with respect to the parameter $d$. This can be accomplished via the implicit function theorem.
In particular, let $z \in \reals^L$ be all the primal and dual variables of~\eqref{eq:dispatch} and define $F(z, d) : \reals^L \times \reals^{NT} \rightarrow \reals^L$ to represent the Karush-Kuhn-Tucker (KKT) conditions of~\eqref{eq:dispatch} parametrized by $d$, where $L = T(4N+4M+7K+1) + 3K$.
Then, $z$ is a solution to~\eqref{eq:dispatch} if and only if $F(z, d) = 0$.
According to the implicit function theorem, under modest regularity conditions, we can locally define a differentiable function $z^*(d) : \reals^{NT} \rightarrow \reals^L$ such that $F(z^*(d), d)=0$.

\vspace{0.5em}

\begin{theorem}[Implicit Function Theorem,~\cite{Dontchev2009-ep}] 
\label{thm:implicit}
    Suppose the solution $z_0$ to~\eqref{eq:dispatch} exists uniquely for $d_0 \in \reals^{N \times T}$, i.e., $F(z_0, d_0) = 0$ and $F(z, d_0) \neq 0$ for $z \neq z_0$. 
    In addition, suppose $F(z, d)$ is twice continuously differentiable in both its arguments.
    Then there exists a function $z^*(d)$ such that,
    \begin{align*}
        F(z^*(d), d) = 0, 
    \end{align*}
    for all $d \in \Omega$, where $d_0 \in \Omega \subset \reals^{N \times T}$.
    Moreover, the function $z^*(d)$ is differentiable with Jacobian,
    \begin{align*}
        Dz^*(d) = -\partial_1 F(z^*(d), d)^{-1} \partial_2 F z^*(z(d), d).
    \end{align*}
\end{theorem}
Based on Theorem~\ref{thm:implicit}, we can compute~\eqref{eq:LMEs_implicit} in one of two ways.
First, we could use \textit{forward-mode differentiation} to construct the full Jacobian $Dz^*(d)$, which requires solving the linear system,
\begin{align}
\label{eq:linear-system}
    \partial_1 F \cdot X = - \partial_2 F,
\end{align}
which has $NT$ right hand sides.
Alternatively, we could use \textit{reverse-mode differentiation}, where we solve the linear system,
\begin{align*}
    \partial_1 F^T \cdot x =  \vectorize(e),
\end{align*}
which has a single right hand side, and then compute $\lambda(d) = -\partial_2 F^T x$.
Reverse-mode differentiation is generally more efficient because it requires solving the factorized system fewer times. 
In both cases, we refer to this methodology as \textit{centralized}, since it requires solving a single linear system involving all the problem data.

\subsection{Computational Complexity}
\label{sec:complexity_centralized}

For notational simplicity, we will write $\bar{N} = N + M + K$ to denote the dimension of the physical network.
The dimension $L$ of the partial Jacobian $\partial_1 F(z^*(d), d)$ scales with $O\left(T(\bar{N}+K)\right)$, since there are $T$ timesteps coupled via $O(K)$ coupling constraints.

Both forward-mode and reverse-mode differentiation require factorizing the matrix $\partial_1 F \in \reals^{L \times L}$, which has cubic complexity $O\left((T(\bar{N}+K))^3\right)$ in the worst case.
However, the matrix $\partial_1 F$ is sparse because of the sparse network structure inherent to power systems and because each timestep depends explicitly only on the neighboring ones.
In this case, a direct factorization can often be computed in $O\left((T(\bar{N}+K))^\beta\right)$ for $1 \leq \beta \leq 3$.\footnote{
    Instead of directly factorizing the matrix, one could also apply an indirect method, such as the conjugate gradient (CG) method~\cite{Shewchuk1994-kl}, to each right hand side.
    In the sparse case, when $\beta \approx 1$, these equate to roughly the same complexity.
}

Afterwards, the factorized system must be solved (e.g., using back substitution) for each right hand side.
Because the factorized system remains sparse, each solve has roughly linear $O\left(T(\bar{N}+K)\right)$ complexity.
This means the total complexity will be $O\left((T(\bar{N}+K))^{\max(2, \beta)}\right)$ for forward-mode and $O\left( (T (\bar{N}+K))^\beta \right)$ for reverse-mode.
We empirically observe $\beta \approx 1$ for Jacobians arising from problems in power systems, making reverse-mode differentiation much faster in practice.
Both methods have complexity that scale with $T(\bar{N}+K)$, which may be prohibitively expensive for large networks over long time horizons.


%% file: body/3.DID.tex
\section{Efficient LMEs via Dual Decomposition}\label{sec:DID}

In this section, we describe an algorithm based on dual decomposition for computing LMEs~\cite{Fuentes_Valenzuela2024-qp} and describe the regimes in which it yields significant computational advantages compared to the centralized method.

\subsection{Dual Decomposition}

In the dispatch problem~\eqref{eq:dispatch}, the constraint \begin{align} \label{eq:soc-evolution}
    s_{k, t+1} = s_{kt} - p_{kt}
\end{align} 
couples the different time periods.
Without this constraint,~\eqref{eq:dispatch} would reduce to $T$ independent single-period optimization problems. 
By applying dual decomposition to~\eqref{eq:dispatch}, we can decouple the subproblems across time periods.
We introduce a variable $\snext \in \reals^{K \times T}$ and replace~\eqref{eq:soc-evolution} with the constraints $\snext = s - p$ and $s_{k, t+1} = \snext_{kt}$ for all $k \in [1:K]$, $t \in [1:T]$. 
The variable $\snext_t$ effectively acts as a local copy of the state of charge at the next timestep.
The constraint coupling all timesteps can then be written as 
\begin{align*}
    H(s, \snext) = \begin{bmatrix}
        \snext_1 - s_2 \\
        \snext_2 - s_3 \\
        \vdots \\
        \snext_T - s_{T+1}
    \end{bmatrix}=0. 
\end{align*}
Let $\nu_{kt}(d) \in \reals$ be the optimal dual variable of constraint $s_{k, t+1} = \snext_{kt}$.
Then dual decomposition divides~\eqref{eq:dispatch} into $T$ optimization problems. 
For timestep $t$, we have
\begin{align} \label{eq:local}
\begin{array}{lll}
    \textrm{minimize} 
    & \sum_{n} c_{nt} g_{nt} - \nu_{t-1}(d)^T s_{t} + \nu_t(d)^T \snext_{t} \\[0.5em]
    \textrm{subject to} 
    & g_t + B p_t - d_t = Af_t \\
    & f_t = \diag(b) A^T \theta_t \\
    & 0 \leq g_t \leq g_t^{\max} \\
    & -f^{\max} \leq f_t \leq f^{\max} \\
    & -p^{\max} \leq p_t \leq p^{\max} \\
    & 0 \leq s_t \leq s^{\max} \\
    & (\theta_{t})_1 = 0, \\
    & \snext_{t} = s_{t} - p_{t}
\end{array}
\end{align}
where the variables are $g_t, \theta_t \in \reals^N$, $f_t \in \reals^M$, and $p_t, s_t, \snext_t \in \reals^K$.\footnote{%
When $t=1$ or $t=T$ the problem will have one fewer term in the objective and an additional constraint for the initial and final condition.
}
Solving~\eqref{eq:local} for every time period $t = 1, \ldots, T$ subject to the constraint $H(s, \snext) = 0$, implicitly enforced through the dual variable $\nu$, gives the solution to the original dispatch problem in~\eqref{eq:dispatch}.

We let $g_t^*(d) : \reals^N \rightarrow \reals^N$ be the solution map of the $t$-th problem.
We now treat $\nu(d)$ as a parameter of the optimization problem.
By the chain rule, the Jacobian $Dg_t^*(d)$ of the solution map of the $t$-th problem is
\begin{align} \label{eq:decomposed-jacobi}
    Dg_t^*(d) = \partial_{d_t} g_t^*(d_t, \nu(d)) + \partial_\nu g_t^*(d_t, \nu(d)) \cdot D\nu(d).
\end{align}
We call $\partial_{d_t} g_t^*(d_t, \nu) \in \reals^{N \times N}$ the \textit{local} Jacobian at time $t$, since it gives the sensitivity of generation at time $t$ to demand at time $t$.
We call $\partial_\nu g_t^*(d_t, \nu(d)) \in \reals^{N \times TK}$ and $D\nu(d) \in \reals^{TK \times TK}$ the \textit{interface} Jacobian and \textit{coupling} Jacobian, respectively.
To calculate both the local and interface Jacobians, we can simply apply Theorem~\ref{thm:implicit} to the KKT conditions of~\eqref{eq:local}.
To calculate the coupling Jacobian, we need to derive an implicit relationship between $\nu$ and $d$.

\subsection{Computing the Coupling Jacobian}

We can decompose the coupling constraint $H$ into $H(s, \snext) = \sum_t h_t(z_t)$, where $z_t \in \reals^{L_t}$ are the primal and dual variables of the $t-$th subproblem defined by~\eqref{eq:local}, and
\begin{align*}
    h_t(z_t) = \begin{bmatrix}
        0 & \cdots & 0 & -s_t & \snext_t & 0 & \cdots & 0
    \end{bmatrix}^T.
\end{align*}
Note that the function $h_t$ only depends on (two of) the variables associated with subproblem $t$.
At optimality,
\begin{align*}
    \sum_t h_t(z_t^*(d_t, \nu(d))) = 0.
\end{align*}
Differentiating with respect to $d$, we obtain
\begin{align*}
    \sum_t 
    D h_t \cdot \left(
    \partial_{d_t} z_t^* + \partial_\nu z_t^* \cdot D \nu
    \right)
    = 0,
\end{align*}
where we have dropped the arguments for ease of notation.
Rearranging terms, we obtain the linear system,
\begin{align} \label{eq:coupling-system}
    \left( \sum_t 
    D h_t \cdot \partial_\nu z_t^* \right) \cdot D \nu
    = 
    -\sum_t D h_t \cdot
    \partial_{d_t} z_t^*,
\end{align}
which yields an expression for calculating $D\nu$.
In summary, we can compute the Jacobian $Dg^*(d)$ in three steps.
\begin{enumerate}
    \item Evaluate the local and interface Jacobians $\partial_{d_t} z_t^*$ and $\partial_\nu z_t^*$ by applying Theorem~\ref{thm:implicit} to the KKT conditions of the local problem~\eqref{eq:local}.
    \item Evaluate the Jacobians $Dh_t$ analytically, then formulate and solve the linear system in~\eqref{eq:coupling-system} to calculate $D\nu$.
    \item Use~\eqref{eq:decomposed-jacobi} to calculate $Dg^*(d)$.
\end{enumerate}

Similarly as for the centralized method in Section~\ref{sec:LMEs}, the decentralized computation of LMEs can either be performed using \textit{forward-mode} or \textit{reverse-mode} differentiation;
the latter method will be faster because it requires solving linear systems with fewer right hand sides.

\subsection{Computational Complexity}\label{sec:complexity_distributed}

The computational cost of the decentralized scheme is dominated by the time to factorize and solve the relevant linear systems.
This can be split into two main components.
First, we evaluate the $T$ local and interface Jacobians.
Evaluating each Jacobian requires solving a linear system of size $O(\bar{N})$, which takes $O(\bar{N}^\beta)$ operations.
Second, we incorporate the coupling term, which requires solving the linear system in~\eqref{eq:coupling-system} of size $O(TK)$, which takes $O\left((T K)^\beta\right)$.
Therefore, the total complexity of the distributed algorithm is $O\left(T\bar{N}^\beta + T^\beta K^\beta\right)$, in comparison with the centralized algorithm which takes $O(T^\beta (\bar{N}+K)^\beta)$.
As a function of the time horizon $T$, the speedup of the algorithm is,
\begin{align} \label{eq:complexity}
\begin{split}
    \eta(T) = \frac{T^\beta (\bar{N}+K)^\beta}{T\bar{N}^\beta + T^\beta K^\beta}.
\end{split}
\end{align}
For small values of $T$, we expect $\eta(T)$ to be increasing as a function of $T$ while for large values of $T$, we expect it to approach a constant. 
The limiting speedup achieved depends on the value of $\beta$. 
Two regimes of interest emerge. 
\paragraph*{$\beta > 1$}
In this case, the speedup is
\begin{equation*}
    \lim_{T \to \infty} \eta(T) = \frac{(\bar{N}+K)^\beta}{K^\beta} > 1.
\end{equation*}
Therefore, one should expect a speedup that is up to cubic, for dense matrices, in $\bar{N}/K + 1$.

\paragraph*{$\beta \approx 1$} 
In power systems problems, both linear systems are sparse and structured which leads to approximately linear complexity, i.e. $\beta \approx 1$.
In this case,
\begin{equation*}
    \lim_{T \to \infty} \eta(T) = 1,
\end{equation*}
i.e., the decentralized differentiation method does not offer any advantage. 
In the following section, we address this limitation by parallelizing the decentralized computation.

\subsection{Parallel Computation}

The decentralized differentiation method involves solving $T$ local linear systems. 
This is an embarrassingly parallel operation;
on a machine with $T$ processors, we can solve all $T$ linear systems independently, requiring just $O(\bar{N}^\beta)$ operations per processor.
Therefore, another advantage of the decentralized algorithm is the ability to parallelize the local linear system solves across multiple processors.
This is similar to how dual decomposition-based optimization algorithms, e.g., the alternating direction method of multipliers~\cite{Boyd2011-xn}, parallelize local variable update rules.

In this case, we can rewrite the speedup in~\eqref{eq:complexity} as
\begin{align} \label{eq:complexity_2}
    \eta(T) = \frac{T^\beta (\bar{N}+K)^\beta}{\bar{N}^\beta + T^\beta K^\beta}.
\end{align}
For small values of $T$, $\eta(T)$ scales as $O(T^\beta)$.
For large values, $\eta(T)$ asymptotically approaches $ (\bar{N}/K + 1)^\beta$. Parallelization therefore enables significant speedups even when $\beta\approx 1$.

The theoretical maximum speedup scales as $O\left((\bar{N}/K + 1)^\beta\right)$. 
Recent data from the U.S.\ EIA report around 450 utility-scale energy storage systems for electricity generation across the U.S.~\cite{US_Energy_Information_Administration_EIA_undated-ek}, and open-source models of the U.S.\ transmission network~\cite{Tehranchi2023-kr} contain around $N=\num[group-separator={,}]{80000}$ nodes. 
In this case, $\bar{N}/K \approx 200$, potentially leading to very large speedups over long time horizons.
Even when other dynamic constraints like generator ramping limits are introduced, they will still constitute a relatively small part of the total dimension of the problem.

%% file: body/4.NumericalExperiments.tex
\section{Numerical Experiments}
\begin{figure*}[t]
    \centering
    \includegraphics[scale=1]{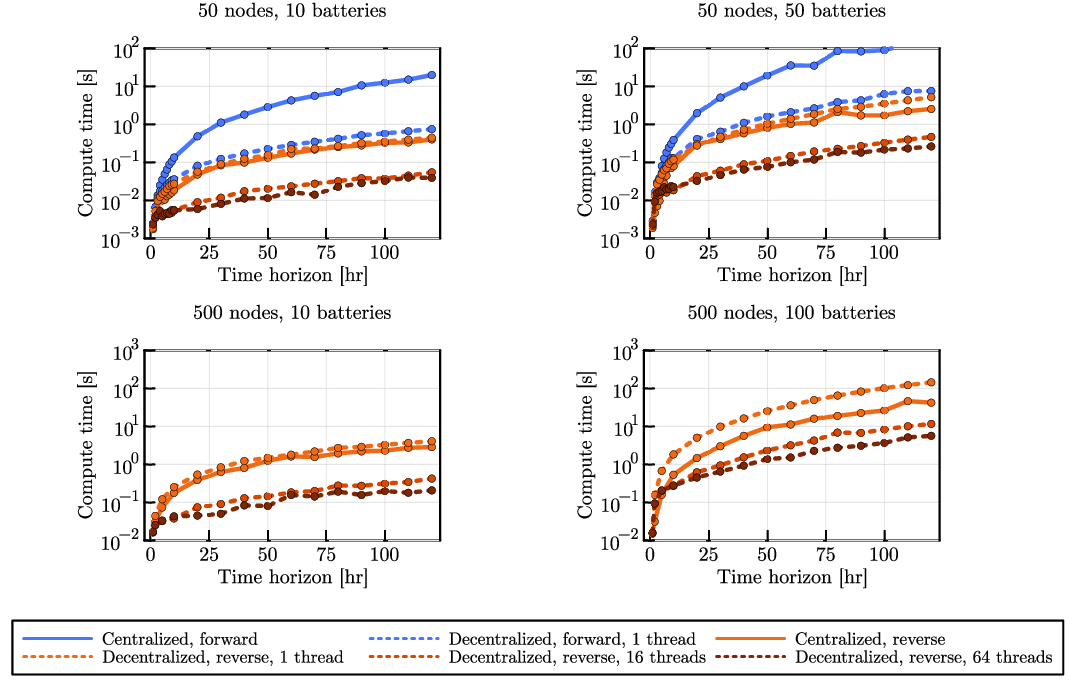}
    \caption{Computation time for both the centralized and decentralized differentiation methods on 50-node and 500-node networks with increasing storage penetration. 
    Forward mode differentiation is much less efficient than reverse mode. 
    Parallel computation enables significant speedup in reverse mode.}
    \label{fig:computation_times}
\end{figure*}

\begin{figure}
    \centering
    \includegraphics[scale=1]{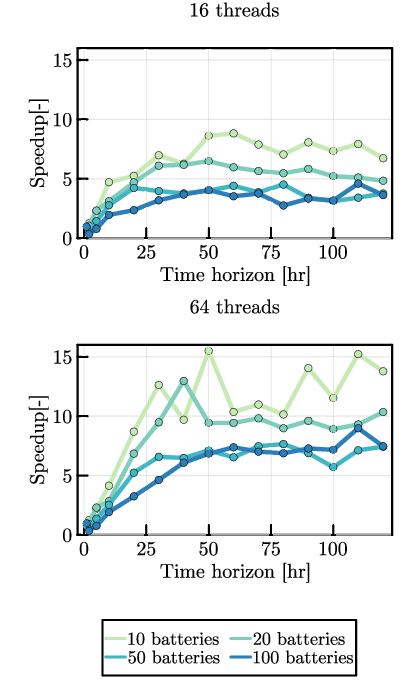}
    \caption{Speedups achieved for different numbers of allocated threads in the 500-node networks. The speedup is defined as the minimum centralized runtime divided by the minimum decentralized runtime. The speedup curves consists of a linear regime followed by a performance plateau, as predicted. The maximum speedup achieved increases with parallelism, and decreases with the number of batteries in the system.}
    \label{fig:speedups}
\end{figure}


In this section, we demonstrate the performance of the proposed method using several network models of the U.S.\ Western Interconnect. 
We empirically validate the computational advantages provided by the decentralized method using both forward-mode and reverse-mode differentiation, with and without parallel computation.

\subsection{Data}

For these experiments, we use the PyPSA USA grid dataset~\cite{Tehranchi2023-kr}, which consists of a large network model of the U.S.\ Western Interconnect that can be dynamically resized.\footnote{
The PyPSA-USA dataset is fully open source and can be found online at \url{https://github.com/PyPSA/pypsa-usa}.
}
The network model is based on the Breakthrough Energy network~\cite{Xu2021-fc} combined with EIA load and generation data.
The nodes in the network can be clustered to between 30 and 4786 nodes as needed;
clustering is performed using nodal renewable profiles.
In our experiments, we solve the dispatch model at an hourly resolution and cluster the network model to either 50 or 500 nodes.
To explore the behavior of the proposed method under different scenarios, we configure the number of batteries and storage penetration of the network for each specific experiment. 
We add batteries to the nodes with the highest peak renewable generation and set the total storage penetration to 10\% of peak demand.
The performance of the algorithm does not depend on the location or capacity of batteries, but only on the number of batteries, as detailed above.


\subsection{Computational Performance}


We empirically compare the computational performance of the different methods. 
We compute LMEs using the centralized and decentralized approaches, with both forward and reverse-mode differentiation, with and without parallelism. 
By repeating such computations on networks of varying sizes and storage penetration, we isolate the main elements that govern the computational performance gains.
All our experiments are run on an AMD EPYC~776 proccessor with 64~cores (128~threads) and 256~GB of memory.
In the results below, each experiment consists of 10 independent trials. The runtime is defined as the minimum across the 10 trials of the total time dedicated to solving linear systems. 
The speedup is defined as the ratio of these minima between the centralized and decentralized method.

Results for the 50-node and 500-node networks are reported in Fig.~\ref{fig:computation_times}. 
First, for the 50-node network, we observe that forward-mode differentiation is significantly more expensive than reverse-mode.
This is not surprising, since forward mode explicitly constructs large Jacobian matrices, while reverse mode only uses vector-Jacobian products. 
Second, in reverse mode, the sequential---i.e., non-parallelized---decentralized differentiation method displays similar computation time as the centralized method. 
Under the complexity model of Section~\ref{sec:complexity_distributed}, this implies a value of $\beta \approx 1$ for these problems.
Parallelization, however, greatly improves the performance of the decentralized method: the computation time steadily decreases with the number of parallel threads.
For the 500-node networks, we omit forward-mode results because they require a prohibitively large amount of memory and compute time to run.
However, we generally observe similar results to that of the 50-node networks: sequential decentralized computation performs similarly (in fact, slightly worse) than the centralized computation, but introducing parallelism enables a significant speedup. 

The speedups (ratio of runtimes) achieved in the 500-node networks are further analyzed in Fig.~\ref{fig:speedups}. 
Consistent with expectations from~\eqref{eq:complexity_2}, the speedup curves display two distinct regimes. 
For short time horizons, the speedup increases with the number of timesteps in the problem; at longer time horizons, it stabilizes around a value that decreases with the number of batteries.
This asymptotic behavior occurs because, for long time horizons, the cost of solving the linear system associated with coupling Jacobian dominates the cost of computing LMEs.
As the number of threads increases, the maximum speedup achieved also increases, reaching up to 15x for 64 threads in the 500-node network with 10 batteries. 
We attribute the observed oscillations to variable compute conditions (e.g., non-deterministic thread scheduling) and finite sample sizes.


%% file: body/5.Conclusion.tex
\section{Conclusion}

Locational marginal emissions (LMEs) are important metrics for understanding the sensitivity of emissions to marginal changes in load in a power network. 
These networks are inherently coupled over time via storage constraints and generator ramping limits.
Accurately modeling these constraints becomes increasingly important with growing intermittent renewable and storage penetration.
However, computing LMEs over long time horizon can become prohibitive for large networks.
In this paper, we address this computational bottleneck through a parallel, reverse-mode implementation of decentralized implicit differentiation.
For problems in power systems, which are generally sparse and highly structured, we show both theoretically and empirically that parallelism is necessary to accelerate differentiation and demonstrate up to a 15x speedup on modestly sized networks.
Finally, we note that our method is general in that it can be used to compute arbitrary power system sensitivities, not just marginal emission rates, and applied to any convex optimization-based dispatch model.

